\documentclass[10pt,article,twocolumn]{IEEEtran}
\usepackage{balance}
\usepackage{cite}
\usepackage{savesym}
\usepackage{amsmath}
\savesymbol{iint}
\restoresymbol{TXF}{iint}
\usepackage{amsfonts,amssymb}
\usepackage{amsmath,bm}
\usepackage{setspace}
\usepackage{epsfig,endnotes}
\usepackage{url}
\usepackage{graphicx}
\usepackage{epstopdf}
\usepackage[pdf]{pstricks}
\usepackage[small]{caption2}
\usepackage{fmtcount}
\usepackage{wasysym}
\usepackage[small]{subfigure}
\usepackage{textcomp}
\usepackage{eso-pic}    
\usepackage{wireless-utils}
\usepackage{xcolor,colortbl}
\allowdisplaybreaks
\newcommand{\tabincell}[2]{\begin{tabular}{@{}#1@{}}#2\end{tabular}}

\begin{document}
\title{\huge{Passive Sensing and Communication Using Visible Light: Taxonomy, Challenges and Opportunities}}
\author{
  \IEEEauthorblockN{Qing Wang and Marco Zuniga} \\
  \IEEEauthorblockA{Delft University of Technology, the Netherlands\\
 	\textit{Email: \{q.wang-5, m.a.zunigazamalloa\}@tudelft.nl}}
}

\maketitle

\begin{abstract}
For more than a century, artificial lighting has served mainly for illumination. Only recently, we start to transform our lighting infrastructure to provide new services such as indoor localization and network connectivity. These innovative advancements rely on two key requirements: the ability to modulate light sources (for data transmission) and the presence of photodetectors on objects (for data reception). But not all lights can be modulated and most objects do not have photodetectors. To overcome these limitations, researchers are developing novel sensing and communication methods that exploit \emph{passive} light sources, such as the sun, and that leverage the external surfaces of objects, such as fingers and car roofs, to create a new generation of cyber-physical systems based on visible light. In this article we propose a taxonomy to analyze these novel contributions. Our taxonomy allows us to identify the overarching principles, challenges and opportunities of this new rising area. 
\end{abstract}


\section{Introduction} \label{sec_intro}
The Internet of Things and Cyber-Physical Systems are enabling a new computing era characterized by a tight integration between the virtual and the real worlds. This new era heavily depends on \emph{wireless} communication and sensing, and these wireless interactions mostly rely on the radio-frequency (RF) band. We argue that in this computing era, the visible light spectrum could play a far greater role than that it is currently playing. To achieve this goal, we must investigate new methods for \emph{passive} sensing and communication.

Visible light is present everywhere and is gaining significant interest as a medium to connect computers and objects. Thanks to advances in visible light communication (VLC), LEDs can now be modulated to transmit data without affecting the illumination perceived by people. Thus we can piggyback wireless communication on top of LED illumination almost for free.
This breakthrough creates a new range of exciting applications. Philips transforms lighting infrastructures to provide localization services~\cite{philipsincarrefour}, PureLiFi provides Internet connectivity through LEDs at data rates comparable to WiFi~\cite{pureLiFi}, and Disney develops a new generation of interactive toys~\cite{ToysVLC}.

\textbf{Limitations of \emph{active} methods.} The above applications are transforming the role of our lighting infrastructures, \emph{but they assume two key requirements: light sources can be modulated to transmit information and objects have photodetectors to receive that information.} These requirements limit how visible light can be exploited for sensing and communication. 

\emph{1) Limitation at the transmitter (TX) side.} Many light sources cannot be modulated. For example, we cannot modulate the sun, but it would be transformative if we could leverage sunlight for wireless communication. Most of the optical radiation in our environments remains largely unused, not only sunlight but also plenty of artificial lighting. Currently, we mainly exploit passive optical radiation to harvest energy, but we should exploit it to convey information as well.

\emph{2) Limitation at the receiver (RX) side.} Most objects do not have photodetectors. It would be equally transformative if we could connect any object using visible light. Furthermore, even objects with photodetectors, such as smartphones with cameras, are only useful when held with line-of-sight (LOS) toward luminaries. This limitation is not present in RF systems, because RF can travel through opaque and solid objects. 

\def \caseA {\textit{full-active}}
\def \caseB {\textit{passive-src}}
\def \caseC {\textit{passive-obj}}
\def \caseD {\textit{full-passive}}

\section{A Taxonomy to Analyze \emph{Passive} Methods}\label{sec_taxonomy} 

To boost using visible light as sensing and communication medium, researchers are investigating passive methods for scenarios with (i) \emph{passive light sources}, which do not modulate information; and (ii) \emph{passive objects}, which do not have photodetectors. These efforts however are loosely connected. \emph{To consolidate this nascent area, it is necessary to have a common framework to identify the general principles, challenges and opportunities of passive methods} for exploiting visible light.

\begin{figure}[t]
	\centering
	\includegraphics[width=1.02\columnwidth]{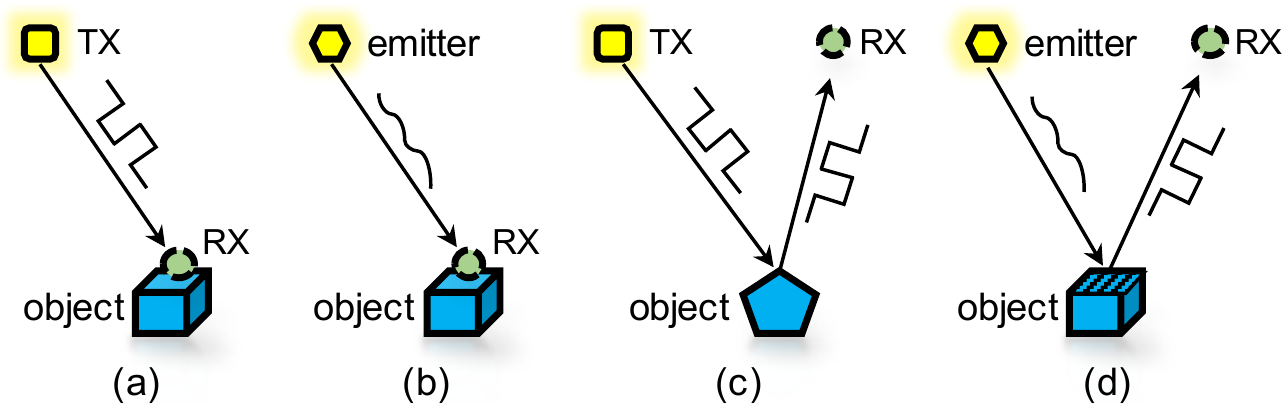}
	\caption{A taxonomy for sensing and communication systems with visible light:
		(a)~\caseA: active source and active object; (b)~\caseB: passive source and active object; (c)~\caseC: active source and passive object; (d)~\caseD: passive source and passive object}
	\vspace{-2mm}
	\label{fig_taxonomy}
\end{figure}

We propose a taxonomy that arranges all efforts related to visible light into four cases. Next, we first describe the traditional scenario (has active light sources and active objects), and then describe unique properties of passive and semi-passive scenarios. Our taxonomy is  illustrated in Fig.~\ref{fig_taxonomy}, where we refer to active sources as \emph{TXs}, and to passive sources as \emph{emitters}.

\emph{\textbf{Case A:} \caseA~(active source, active object)}. This is the most popular application of visible light communication. The goal is to transmit information from a light source to an object. This goal is simple to attain because the light source can be modulated at high frequencies, and the object can decode this data reliably thanks to having a photodetector with LOS toward the luminaries (minimal channel distortions). 

\emph{\textbf{Case B:} \caseB~(\textbf{passive source}, active object)}. Passive light sources change the problem fundamentally. Since we cannot modulate them as in Case A, the goal is \emph{not} to transmit information from the light fixture to the object. The goal now is for the object to \textit{get information about the environment} by measuring \textit{uncontrolled} changes in illumination. The information can still be measured directly by the object because it has a photodetector, but the amount of information is limited and depends solely on the dynamics of the scenario at hand. 

\emph{\textbf{Case C:} \caseC~(active source, \textbf{passive object})}. Passive objects also change the problem fundamentally. Objects can no longer get information about the environment as in Case B, because they have no photodetectors. With passive objects, the photodetectors need to be placed in the environment. Thus, the goal changes: instead of having objects getting information about the environment, now the environment \textit{gets information about objects}. Having active sources means that we can send fine grained pulses to get information about objects, but these pulses will be reflected by the objects' external surfaces, and thus, the received signals will be noisy. There is no-line-of-sight (NLOS) between light fixtures and RXs. 

\emph{\textbf{Case D:} \caseD~(\textbf{passive source}, \textbf{passive object})}. This scenario is the most complicated. The presence of passive objects (photodetectors in the environment not on the object), means that the goal is the same as in Case C: \textit{get information about the object.} However, we don't have an active source to modulate information. Thus, the only source of information is the reflection caused by the object's surface. This scenario leads to a compound problem: a noisy generation of information, because there is no active source; and a noisy reception of information, because the signals are reflected (NLOS).

In this paper, we refer to the cases \caseB, \caseC, and \caseD ~as \emph{\textbf{passive}} systems. Furthermore, we classify them into two groups based on their main objective: sensing or communication. We will first focus our discussion on \textbf{\emph{visible light sensing (VLS)}} and then on \textbf{\emph{communication (VLC)}}.
\section{Passive VLS: Architecture and Applications} \label{sec_sensing} 

\subsection{System Architecture} \label{sec_sensing_properties}
The architecture of typically passive VLS systems mainly  consist of three elements: {\it light source, object,} and {\it receiver (RX).} 
The {\it light source} can be anything: an LED (can modulate information), an incandescent bulb (cannot modulate information) or natural light sources such as the sun (uncontrollable). The {\it objects} can be of any form: people, cars, fingers, to name a few. The \textit{RX} is a tiny box containing simple photodetectors, such as photodiodes or cameras.

\subsection{Applications} \label{sec_app}
Passive VLS is enabling many applications. Below, we describe these applications based on our taxonomy (Table~\ref{table_sensing_app} summarizes our classification). 

\emph{Case B: \caseB.} If the RX is placed on the object, Fig.~\ref{fig_taxonomy}(b), the information is obtained from default changes in the light intensity of emitters. For example, LiTell~\cite{LiTell2016} provides sub-meter indoor localization using standard fluorescent lights as emitters and smartphone cameras as RXs. LiTell measures the specific frequency of nearby fluorescent bulbs, and determines the location of the user (camera) after comparing the captured light frequency to those in a database.

\emph{Case C: \caseC.} If the RXs are placed in the environment and the light source can be modulated, Fig.~\ref{fig_taxonomy}(c), the information is obtained from the reflections over the objects' surface. Research studies show that many objects can be monitored with this type of architecture: fingers, cars and people. In Okuli~\cite{Okuli}, the goal is to track a moving finger over a pad. A small LED (active source) and two photodiodes are placed at one side of the pad, and the system maps the location of the finger based on its reflected light intensity measured at the photodiodes. Cars could also be tracked with a similar approach, as shown in~\cite{Zhang2016PassiveLoc}. If street luminaries could modulate their light intensities and have photodiodes co-located with them, the reflections from the cars' surface can be exploited to pinpoint their location accurately. Passive VLS systems are also being used to monitor people~\cite{Li_2015}. In this case the RXs are embedded in the floor and ceiling luminaries send modulated signals. Based on the distortions measured at the RXs, the system can reconstruct a person's posture. 

\emph{Case D: \caseD.} Like the previous case, in this system, the information comes from reflections. But the information is less accurate because the system does not use modulated light sources, Fig.~\ref{fig_taxonomy}(d). Still, researchers are developing interesting applications leveraging people and hands as passive objects. CeilingSee can estimate the occupancy of rooms using ceiling luminaries that also act as receivers~\cite{CeilingSee}. The level of illuminance distortions perceived at the ceiling indicates the number of people present in a room. With a similar approach, we could track a single person within a room by deploying a grid of receivers on the ceiling ~\cite{Ibrahim2016}; and the same tracking goal can be obtained by placing receivers on the floor to measure the shadows cast by people~\cite{LocaLight2016}. Besides monitoring people, passive VLS can also be used to monitor hand gestures. GestureLite exploits ambient light and an array of receivers placed on a wall to recognize hand gestures~\cite{GestureLite}. Based on the light disturbances caused by hand movements in front of the receivers, the system can distinguish ten hand gestures.

\definecolor{myblue}{RGB}{117,182,221}
\begin{table}[t]
	\renewcommand{\arraystretch}{1.05}
	
	\centering
	\caption {\emph{State-of-the-art} passive VLS applications} \label{table_sensing_app}
	\vspace{1mm}
	\resizebox{1.02\columnwidth}{!}{
		\begin{tabular}{| r | l | l |} 
			\hline
			\tabincell{l}{ } & \textbf{\hspace{0mm}  Passive light source}  & \textbf{\hspace{0mm} Active light source}  \\ \hline
			
			\tabincell{r}{\textbf{Passive object}}  & \cellcolor{myblue}  \tabincell{l}{- LocaLight\cite{LocaLight2016} \\ - CeilingSee\cite{CeilingSee}  \\ - GestureLite\cite{GestureLite} \\ - CeilingSensing\cite{Ibrahim2016}
			}  &  \cellcolor{myblue}  \tabincell{l}{- Passive localization~\cite{Zhang2016PassiveLoc} \\ - Human sensing~\cite{Li_2015} \\ - Okuli~\cite{Okuli} } \\ \hline
			
			\tabincell{r}{\textbf{Active object}}  & \cellcolor{myblue} {- LiTell~\cite{LiTell2016}} & \tabincell{l}{- Indoor localization~\cite{Epsilon} } \\ \hline
		\end{tabular} 
	}
\end{table}

\section{Passive VLS: Challenges} \label{sec_sensing_challenges}
The design of passive VLS systems requires tackling unique challenges. Based on our taxonomy and analysis of state-of-the-art, below we describe the three most important challenges. 

\begin{figure}[t]
	\centering
	\includegraphics[width=1.02\columnwidth]{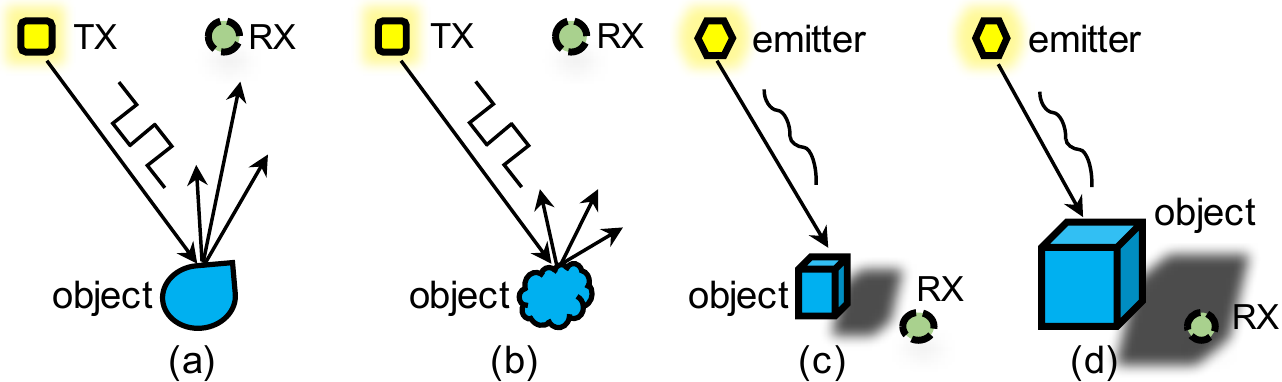}
	\caption{An illustration of \emph{Challenge 1}. There is no {\it `standard'} object to be sensed. The object's shape and reflection properties determine the direction and intensity of the reflected light, \emph{(a-b)}; and its shape determines the amount of light blocked over the floor, see \emph{(c-d)}.}
	\label{fig_challenge_1}
\end{figure}

\emph{\textbf{Challenge 1:}} \emph{No control over the objects' shape, implies no one-size-fits-all solutions.} In \textit{active} scenarios, it is customary to have a \textit{single} modulation method to communicate with \textit{any} object, because objects carry photodetectors.  However, in scenarios where the object carries no photodetector (cases C and D), we rely on the object's external surface for sensing. In these scenarios, the design and performance of the system is largely determined by the object's properties, but objects have different shapes, sizes and reflection coefficients, as shown in Fig.~\ref{fig_challenge_1}. The object's shape determines the direction of reflected light; its size determines the amount of light that is blocked; and its reflection coefficient, or even cleanness, heavily affect how much light is reflected towards the RX. There is no `standard' object to be sensed. 
Thus, \textit{before designing a passive VLS system, it is central to gain a deep understanding about the object at hand, to design a tailored system}. It is challenging to design one-size-fits-all solutions that can monitor accurately various types of objects.

\emph{\textbf{Challenge 2:}} \emph{No control over the emitters, requires designing more flexible and robust methods for reception.} In scenarios with \emph{active} light sources, RXs are designed to focus on the specific range of frequencies and intensity-levels modulated by the active light fixtures. The effect of other (passive) light sources is filtered out via hardware or software. Passive VLS systems, on the other hand, cannot filter out these passive light sources because \emph{it relies on them for sensing} (cases B and D). But we cannot control the intensity, location or any other property of emitters. Thus, receivers in passive VLS need to work well under a wider range of optical frequencies and intensities. Furthermore, similar to Challenge 1, where the lack of control over objects requires a deeper understanding of reflections; in this case, the lack of control over emitters requires a deeper understanding of the expected illumination conditions (to \textit{fine-tune} the design of algorithms). Overall, loosing the ability to modulate (control) a signal in passive VLS, creates challenges that can only be tackled with a more flexible and robust design at the reception end.

\emph{\textbf{Challenge 3:}} \emph{Monitoring passive objects, requires a high-density of receivers.} In scenarios where photodetectors are placed on top of objects, the RX moves along with the object, and thus, can provide continuous sensing. In scenarios with passive objects, the RXs are fixed to certain places and can only provide information when objects move under their limited field-of-view (FoV). To cover a large sensing area and/or provide fine-grained results, more RXs are indispensable. These denser deployments require not only a careful analysis to minimize the number of RXs, while still guaranteeing a minimum performance level, but they also require designing more energy-efficient RXs to minimize the overall energy footprint of passive VLS systems.

\begin{figure*}[t]
	\centering
	\subfigure[Divide a TX into sub-TXs]
	{\includegraphics[width=0.52\columnwidth]{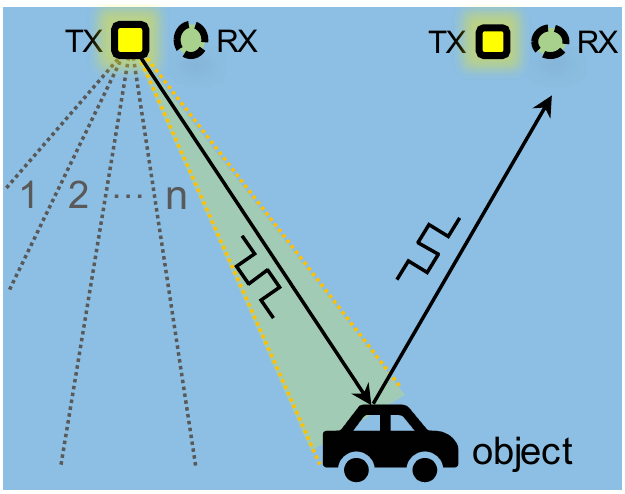}} 
	\hfill
	\subfigure[Map the research opportunities to the research challenges]
	{\includegraphics[width=1.5\columnwidth]{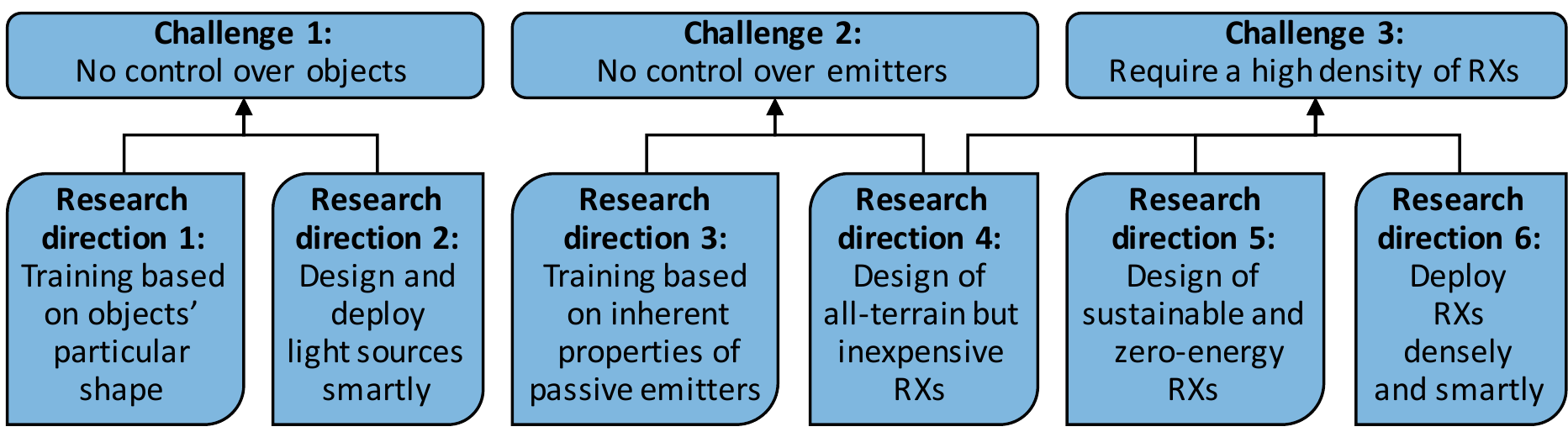}}
	\caption{Future research opportunities in the passive VLS systems}
	\label{fig_sensing_solution}
\end{figure*}

\section{Passive VLS: Research Opportunities} \label{sec_sensing_potentials}

As discussed in the previous section, passive VLS systems expose unique challenges. Below, we describe the new research directions followed by the community to tackle these challenges. Fig.~\ref{fig_sensing_solution}(b) summarizes our findings by mapping challenges to research directions.

\textit{\textbf{Research Direction 1:} Train the system based on the particular shape of the object.} This approach is applied to Cases C and D, where the object is passive. To cope with the unique reflecting properties of different objects (Challenge 1), researchers first create a training database. For example, in CeilingSee~\cite{CeilingSee}, the authors analyze the correlations between the number of people present in a room and the subtle light distortions they cause. These correlations are then used to estimate occupancy levels in real-time. In a different study, GestureLight relies on a test user to create a database of ten hand gestures under different ambient light conditions~\cite{GestureLite}. The database is later used as reference for real-time gesture detection based on machine learning techniques. These two examples work on scenarios with passive lights (Case D). But scenarios with active lights (Case C) can also benefit from a training phase. Okuli~\cite{Okuli} exploits the fact that fingers have circular shapes and reflect light uniformly, to create a database mapping modulated RSS with 2D locations on a pad. And humans or human parts (fingers, hands) are not the only objects that can exploit their unique reflecting properties. In~\cite{Zhang2016PassiveLoc}, the authors leverage the multiple reflecting angles of cars (roof, front/rear glass shield, etc) to provide accurate localization. The system has a database with the cars' angles, and performs simple geometric calculations to identify its location based on the reflected signals. 

The main limitation of these methods is the overhead of creating and maintaining training sets. There are  research opportunities to reduce this training overhead (or even better, to remove it), without affecting the sensing performance greatly. 

\textit{\textbf{Research Direction 2:} Design and deploy light sources in a smarter manner.} This approach is applied to Case C, where the light is active but the object is passive. To compensate for the lack of control over the passive object (Challenge 1), the design and deployment of \textit{active} lights can be \textit{tailored} to improve the performance of the system. For example, in~\cite{Zhang2016PassiveLoc} the authors take advantage of the fact that light fixtures do not have a single LED substrate, but multiple substrates. Instead of modulating all the substrates simultaneously, as in traditional VLC, the authors modulate each substrate individually, which permits a fine-grained localization of passive objects as shown in Fig.~\ref{fig_sensing_solution}(a). If traditional VLC lights are used, i.e. if all LED substrates are modulated simultaneously, a carefully planned deployment can provide rich information about the object. For instance, a \textit{cross-like} deployment of active ceiling lights is used to monitor human postures~\cite{Li_2015}. And not only can standard luminaries benefit from smart design and deployment. In Okuli~\cite{Okuli}, a simple LED light is mounted on a custom-made mechanical structure to control the light reflected by a finger on a tracking pad. 

There are two important aspects to be considered with these research approaches: (i) the overhead and costs associated with modified light fixtures, and (ii) the balance between sensing and illumination. Contrary to RF systems where changes in output powers are not perceived by users, the changes of lighting infrastructure must not affect user experience.

\textit{\textbf{Research Direction 3:} Train the system based on the inherent properties of some passive emitters.} This approach is applied to Case B, where the light source is passive, but the object carries a photodetector. Some light sources have \textit{inherent} properties that can be sensed by the object's photodetector to improve the accuracy of passive VLS (Challenge 2). LiTell uses this approach to achieve sub-meter indoor localization based on unmodified (passive) fluorescent bulbs~\cite{LiTell2016}. First, photodiodes are used to measure the specific frequency of each fluorescent bulb in an indoor space. These measurements are stored in a database. Later, users with smartphone cameras capture the frequency of nearby fluorescent bulbs and look for a match in the database. 

While training can enable/boost the sensing performance of some systems, it comes at the cost of increased overhead (similar drawbacks to Research Direction 1). For example, the characteristic frequency of fluorescent bulbs depends on the surrounding temperature, which means that LiTell may require frequent training updates.
Still, the idea of exploiting the inherent properties of passive sources is an exciting direction that has not been explored much (as seen in Table~\ref{table_sensing_app}). Many research opportunities are present at the intersection of exploiting these inherent properties while reducing (or removing) training overheads.

\textit{\textbf{Research Direction 4:} Design of all-terrain but inexpensive receivers.} This approach is applied to Cases B and D, where light sources are out of the system's control. The receiver thus needs to cope with a wide range of optical frequencies and intensities (Challenge 2). There are receivers with advanced optical filters that can work in various environments, but they are expensive (upwards of 300 USD, such as the Thorlabs-PDA100A). Given that some passive VLS scenarios will require a high density of RXs (Challenge 3), researchers are looking for ways to design \emph{inexpensive} but all-terrain RXs. One such approach is to use LEDs for the dual purpose of emission and reception of light. CeilingSee~\cite{CeilingSee} for example modifies standard luminaries to emit light but also to sense light. In this way the lighting infrastructure is used not only to provide illumination but also to monitor occupancy. Another approach is to combine simple photodetectors, with different receiving characteristics, into a single RX. For instance, PassiveVLC~\cite{PassiveVLC} combines a photodiode and an LED acting as a receiver. The photodiode works well under low illumination conditions, but saturates under high illumination. The LED has the opposite trade-off. The authors use this dual RX mainly for passive communication, which is explained in the next section, but some of their experiments include passive sensing, such as detecting the type of cars present in a parking lot by analyzing the unique optical signatures reflected by the cars' surfaces. 

There are not many efforts aiming at the design of new RXs suited for passive sensing. Most of the existing works inherent RXs used in traditional VLC. There are thus many research opportunities in this direction.

\textit{\textbf{Research Direction 5:} Design of sustainable and zero-energy receivers.} Requiring a high number of RXs to increase the coverage of passive systems, implies a bigger energy footprint (Challenge 3). To sustain the development of passive VLS, it is necessary to design zero-energy cost receivers: RXs should obtain all their energy via harvesting methods. One such approach is followed in LocaLight~\cite{LocaLight2016}, where RXs are deployed over the floor and powered wirelessly using RF. The receivers pinpoint the location of people based on the shadows they cast over the floor. A key problem faced by zero-energy platforms is the trade-off between energy consumption and operational time. Relying on energy harvesting usually implies intermittent operation. For example, Localight only detects objects 50\% of the time due to the limited harvested energy. 

In general, an important research opportunity for passive VLS is to leverage light as a means for communication and energy. RXs could be designed by default with solar panels, together with complementary harvesting methods for periods with low illumination. And the sensing and data processing methods running on these RXs should be designed to be energy efficient from inception.

\textit{\textbf{Research Direction 6:} Deploy receivers densely and smartly.}
This approach is applied to Cases C and D. Due to the limited FoV of most photodetectors, RXs need to be deployed in higher numbers or be carefully deployed to provide the necessary coverage (Challenge 3).
Many studies follow a high-density approach. For example, to monitor people's posture, 324 RXs are deployed in an area of 3$\times$3m. This high density enables the required granularity to track the movements of limbs~\cite{Li_2015}. Similarly, GestureLite adopts a 3$\times$3 RX array over a small space to achieve a recognition accuracy of $98\%$ for ten different hand-gestures~\cite{GestureLite}; and LocaLight deploys dense RXs on the floor to track the location of people by measuring their shadows~\cite{LocaLight2016}. Other studies follow a careful-deployment approach. Okuli places two photodiodes in a custom-designed enclosure to filter out undesirable light while monitoring the location of a finger on a pad~\cite{Okuli}. Another example is shown in~\cite{Ibrahim2016}, where the authors place RXs in suitable locations to track specific events such as the state of doors (open/closed). 

Deploying a high number of RXs is a relatively simple solution to increase coverage, but increases costs across many dimensions: energy, data processing and infrastructure. A more elegant approach, which also opens more research opportunities, is to design a careful placement of fewer RXs.

\section{Beyond Sensing:\\Passive Communication with Visible Light} \label{sec_comm}

Passive communication is more complex than passive sensing, because communication requires sending bits. Thus, to achieve passive communication we need to find ways to modulate visible light without having control over the emitter.

\begin{figure}[t]
	\centering
	\includegraphics[width=.55\columnwidth]{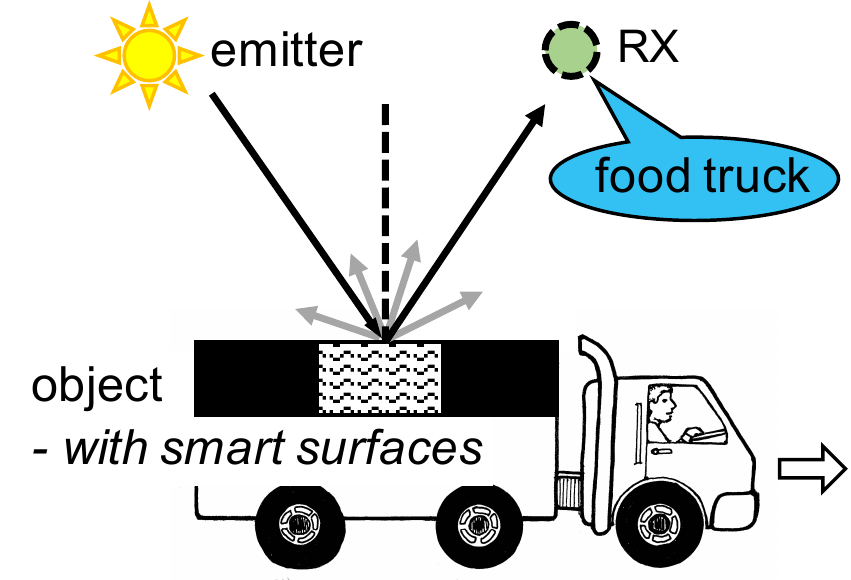}
	\caption{An illustrated application of passive VLC}
	\label{passive_VLC_app}
\end{figure}

The overarching vision of passive VLC is to have objects sense and process data, but instead of communicating this information actively, e.g. via a radio module, objects will adapt the reflective properties of their external surfaces according to the information they want to convey (like a chameleon). In this way, light waves impinging over the smart surfaces will create distinctive patterns, and photodetectors deployed in the surroundings will decode the reflected signals.  A conceptual application of passive VLC is shown in Fig.~\ref{passive_VLC_app}.

\subsection{Architecture and Applications} \label{sec_comm_properties}
 The architecture of a passive communication system is similar to those of passive sensing (Sec.~\ref{sec_sensing_properties}). There are \emph{emitters, objects and RXs.} But there are two key differences: one on the object, the other on the RX. 

\textit{\textbf{Architectural Difference 1:} Objects are covered with smart surfaces.} Passive sensing exploits the \emph{default} external surface of objects. But as described in Challenge 1, objects have different shapes and materials, making them unsuitable to modulate binary information via reflections. To attain some control over reflections, passive VLC systems cover objects with smart surfaces. These surfaces adjust their reflective properties between two states to send information: a high (low) reflective coefficient to transmit a logical one (zero).

\textit{\textbf{Architectural Difference 2:} RXs do not contain cameras.} Some passive sensing scenarios exploit the presence of cameras in smartphones. But if RXs are to be deployed in high densities and in a sustainable manner, cameras are not a suitable option. Cameras are costly, consume more energy and raise up privacy issues. In passive communication the RX is assumed to be always a simple photodiode and/or an LED.

Given the more complex nature of passive communication, researchers have only make some inroads into the problem. There are two main pieces of work in the literature: Retro-VLC~\cite{Retro-VLC} and Passive-VLC~\cite{PassiveVLC}, as summarized in Table~\ref{table_comm_app}. Retro-VLC creates a bidirectional link between an active light source and a surface (object). Both of these elements have a photodiode. For the downlink, the active light transmits information to the surface with traditional VLC. For the uplink, the surface replies back by using an LCD shutter to absorb and reflect the impinging light. Retro-VLC is designed for static objects. Passive-VLC~\cite{PassiveVLC}, on the other hand, is a fully passive communication system for mobile objects. It leverages sunlight to transmit information. An RX is placed on a pole in parking lot. The objects are cars, whose roofs are covered with barcodes consisting of materials with different reflective properties. As the cars pass by, the sunlight impinging on their roofs is modulated by the barcodes and decoded by the RX.

\definecolor{myblue}{RGB}{117,182,221}
\begin{table}[t]
	\renewcommand{\arraystretch}{1.05}
	
	\centering
	\caption {\emph{State-of-the-art} passive VLC applications} \label{table_comm_app}
	\vspace{1mm}
	\resizebox{\columnwidth}{!}{
		\begin{tabular}{| r | l | l |} 
			\hline
			\tabincell{l}{ } & \textbf{\hspace{0mm}  Passive light source} & \textbf{\hspace{0mm} Active light source}  \\ \hline
			
			\tabincell{r}{\textbf{Passive object}}  & \cellcolor{myblue}  {- Passive-VLC~\cite{PassiveVLC} }  &  \cellcolor{myblue} {- Retro-VLC~\cite{Retro-VLC}} \\ \hline
			\tabincell{r}{\textbf{Active object}} & \cellcolor{myblue}  & \tabincell{l}{- Traditional VLC~\cite{pureLiFi} } \\ \hline
		\end{tabular} 
	}
\end{table}

\subsection{Challenges} \label{sec_comm_challenges}

Passive VLC creates a new wireless channel that inherits many of the challenges encountered in passive sensing. Below, we first describe the similarities and differences with the three challenges mentioned in Section~\ref{sec_sensing_challenges}, and then, we introduce new challenges that pertain only to passive VLC.

\textit{Differences with \textbf{Challenge 1}.} Compared to passive sensing that exploits the default surface of objects, passive communication covers objects with surfaces having distinctive reflective properties. Thus, the reflections are not as random as those observed in passive sensing. This ability to control reflections is key to modulate binary information.
 
\textit{Similarities with \textbf{Challenge 2}.} Similar to passive sensing, passive communication also relies on uncontrolled light sources. Thus, RXs need to be flexible and robust to operate in a wider range of illumination conditions.

\textit{Similarities with \textbf{Challenge 3}.} Passive sensing and communication share the same coverage problem. To collect as much information as possible from \textit{passive} objects, many RXs need to be deployed and they have to be energy efficient to reduce the overall energy footprint of the system. 

\textit{\textbf{Challenge 4:} Modulation of passive light requires control over reflections.} Passive VLC heavily depends on the ability to modulate reflections. Due to this reason, we need to design smart surfaces with three key characteristics: high mutability, fine granularity and high energy efficiency. High mutability is required to change rapidly (in the order of ms or less) the reflective properties of the surface between high and low reflection states. Fine granularity is required to control the symbol width of barcodes to encode as much information as possible over the surfaces. A high mutability and granularity would increase the throughput of the system. Finally, the surface should not require high amounts of energy to achieve the required mutability and granularity. Many objects do not have connections to batteries or power outlets, and the energy required to control their surfaces may be obtained only from light itself (harvested through solar panels).

\textit{\textbf{Challenge 5:} The object determines the encoding of information, but we have no control over the object.} Compared to existing communication systems, passive VLC faces unique challenges due to the lack of TXs. In traditional systems, the TX controls the packet size and a symbol's period. In passive communication, these parameters depend on the object's size and speed, and these dependencies cause two problems. \textit{First}, the object's size limits the amount of information that can be encoded. Symbols cannot be too narrow, else they may not be detected; but they cannot be too broad either, else too little information is encoded. It is thus essential to estimate the optimal (minimum) symbol width to maximize the channel's throughput. \textit{Second}, changes in the object's speed can distort symbols' periods. Consider a  Low-High-Low symbol sequence on top of an object moved by a person. The person could walk faster, slower or pause at any point in time. These dynamics would change the duration of symbols, leading to many possible decoding outcomes: LHHL, LHL, LHHLL, etc. In traditional systems the symbol duration within a packet changes minimally, and different methods have been devised to cope with small drifts. Passive communication requires new decoding methods to overcome the larger variations caused by variable objects' speed. 

\balance

\subsection{Research Opportunities} \label{sec_comm_potentials}
Passive VLC is still in its infancy. Below we describe the progress made thus far by the community and the research opportunities we foresee. 

\textit{\textbf{Research Direction 7:} Analysis of smart surfaces.} The design of smart surfaces requires a thorough analysis of various materials. Thus far, researchers have only evaluated basic materials such as aluminum foil and black cardboard~\cite{PassiveVLC}, or screen-based solutions such as LCD shutters~\cite{Retro-VLC}. There are however other materials that could be used such as smart glasses or microblinds. These smart materials are being developed to control the amount of sun radiation in buildings, but they could be used to modulate information as well. We need a thorough understanding about the performance of these materials based on the three metrics mentioned in Challenge 4: mutability, granularity and energy efficiency.

\textit{\textbf{Research Direction 8:} Design of novel decoding methods.} There are no solutions for the research problems introduced in Challenge 5: finding optimal symbol widths to maximize throughput and designing novel encoding methods to cope with variable speeds. In Passive-VLC~\cite{PassiveVLC}, the authors drive cars at constant speed and do not provide insights about the maximum throughput that the system can achieve. Furthermore, the experiments did not consider `collisions' (two objects passing under the same FoV simultaneously) or signal distortions due to damages or dirt on the surfaces. In general, passive VLC is a new area with plenty of opportunities for novel contributions.
   
\section{Conclusion} \label{sec_conclusion}

Considering the increasing attention that visible light is getting as a medium for sensing and communication, in this paper we introduced a taxonomy to classify various passive methods. Our taxonomy allowed us to identify five macro challenges and eight general research directions in this nascent area. Our analysis shows that, by and large, the focus of the community is on applications for monitoring passive objects (objects without photodetectors). Researchers are proposing novel methods to monitor people, fingers, hands and cars. There is less research activity on the more complex problem of passive communication, where the main challenge is to modulate visible light without having control over the light source. We envision that in the future, passive sensing and communication with visible light will enable a new generation of cyber-physical systems, one that will connect \textit{everyday} objects with the vast number of \textit{passive} light sources in our environments.
\bibliographystyle{IEEEtran}
\bibliography{ref}
\end{document}